\documentclass[usenatbib]{aa}

\usepackage{graphicx}
\usepackage{amssymb}
\usepackage{epsfig}
\usepackage{natbib}
\bibpunct{(}{)}{;}{a}{}{,}

\def\pl{{\sc pl}}
\def\compps{{\sc compps}}

\def\chiq{$\chi^2$}

\def\nh{ {$N_{\rm H}$}}
\def\dbb{ {\sc diskbb}}

\begin{document}

\title{INTEGRAL spectroscopy of the accreting millisecond pulsar XTE
  J1807--294 in outburst}
\author{M. Falanga\inst{1}\fnmsep\thanks{\email{mfalanga@cea.fr}},
J. M. Bonnet-Bidaud\inst{1},
J. Poutanen\inst{2},
R. Farinelli\inst{3},
A. Martocchia\inst{4},
P. Goldoni\inst{1},
J. L. Qu\inst{5},
L. Kuiper\inst{6},
A. Goldwurm\inst{1}}

\offprints{M. Falanga}
\titlerunning{Broad-band spectrum of XTE J1807--294}
\authorrunning{M. Falanga et al.}

\institute{CEA Saclay, DSM/DAPNIA/Service d'Astrophysique (CNRS FRE
  2591), F-91191, Gif sur Yvette, France
\and  Astronomy Division, P.O.Box 3000, FIN-90014 University of
  Oulu, Finland
\and Dipartimento di Fisica, Universit\`a di Ferrara, via Paradiso
  12, 44100 Ferrara, Italy
\and Observatoire Astronomique, 11 Rue de l'Universit\'e, F-67000
Strasbourg, France
\and Laboratory for Particle Astrophysics, Institute of High Energy
Physics, CAS, Beijing 100039, P. R. China
\and SRON National Institute for Space Research, Sorbonnelaan 2,
3584 CA Utrecht, The Netherlands
 }

\abstract{
The fourth transient accreting millisecond pulsar
XTE J1807--294 was observed during its February/March 2003 outburst by
{\em INTEGRAL}, partly simultaneously with the {\em XMM-Newton} and
{\em RXTE}
satellites. We present here the first study of the 0.5--200 keV broad-band
spectra of the source.
On February 28, the source spectrum was   consistent with thermal
Comptonization by electrons of temperature $\sim 40$ keV, considerably
larger than the value ($\sim 10$ keV) previously derived from the low
energy {\em XMM-Newton} data alone.
The source is detected by {\em INTEGRAL} up to 200 keV with a
luminosity in the energy band (0.1--200) keV of
  $1.3 \times 10^{37}$ erg s$^{-1}$ (assuming a distance of 8 kpc).
22 days later the luminosity dropped to  $3.6 \times 10^{36}$ erg s$^{-1}$.
A re-analysis of {\em XMM-Newton} data yields the orbital Doppler
variations of the pulse period and refines the previous ephemeris.
For this source having shortest orbital period of any known binary
radio or X-ray millisecond pulsar, we constrain the companion mass
$M_{\rm c} < 0.022 $M$_{\odot}$, assuming minimum mass transfer driven
by gravitational radiation. Only evolved dwarfs with a C/O composition
are consistent with the Roche lobe and gravitational radiation
constraints, while He dwarfs require an unlikely low inclination.
\keywords{accretion, accretion disks -- binaries: close -- stars:
individual (XTE J1807--294) -- stars: neutron} }
\maketitle

\section{Introduction}
\label{sec:intro}

Neutron stars (NS) in low-mass X-ray binaries (LMXB)
are spun up by accretion to millisecond periods.
The   end point of their evolution  is expected to be a
millisecond pulsar (MSP), i.e. a rapidly spinning NS with a rather weak,
$\sim 10^{8}$ G, surface magnetic field (for a review see \citealt*{bh91}).
Although evidences for  rapidly spinning neutron stars in LMXBs
were obtained from the burst oscillations that were seen during type I
X-ray bursts in several systems \citep[see][]{s01} and from the discovery
of kHz QPOs \citep[see][for a review]{vdk00},
the detection of millisecond pulsations in the
persistent emission remained  elusive for many years until the
discovery of the first accreting millisecond pulsar by \citet{wk98}.
Since that time, a total of
six accreting MSPs have been detected (see review by  \citealt*{w05}).
The last one, IGR J00291+5934, was discovered very recently
with {\it INTEGRAL} \citep{e04} and turned out to be the fastest
rotating object among these six with the frequency of 599 Hz \citep{mss04}.
Other pulsars' spin frequencies are between 180 and 435 Hz.
All of the accreting MSPs are X--ray transients. The accreting
  MSPs spend most of the time in a quiescent state, with X--ray
 luminosities of order  
of $10^{31}-10^{32}$ erg s$^{-1}$ and sometimes go into outbursts
reaching luminosities of $\le 10^{38}$ erg s$^{-1}$.
The millisecond pulsar XTE J1807--294 was discovered in outburst on
February 2003 with the {\it Rossi X--ray Timing Explorer} ({\it RXTE})
during the monitoring observations of the Galactic
  center region \citep{m03}.  It has
a pulse period of 5.25 ms. The orbital period of $\sim$ 40 min
is the shortest one of any known binary radio or X-ray millisecond
pulsar.
An {\em XMM-Newton} Target of Opportunity observation have been used
to derive the low energy spectrum and the source pulse profile
\citep{cam03,k04}. In this paper, we analyze the {\it INTEGRAL} hard
X-ray spectra of this source obtained simultaneously with {\it
  XMM-Newton} and {\it RXTE} during the February/March 2003 outburst.

\section{Observations and data analysis}

\subsection{INTEGRAL}
\label{sec:integral}

The pulsar was observed by {\em INTEGRAL}  \citep{w03}
twice. The first observation was obtained during
a Target of Opportunity (ToO) pointing of the Galactic center region,
performed from February 28 to March 1, 2003 (satellite revolution
46) which was contemporaneous with the {\em RXTE}
observation of the pulsar. The second observation during a Galactic
Center Deep 
Exposure (GCDE) observation was contemporaneous with both the {\em XMM-Newton}
and {\em RXTE} observations on March 20--22, 2003 (satellite
revolutions 52/53). Hereafter we call the two datasets rev-46 and rev-52,
respectively.

We observed XTE J1807--294  with one of the main gamma-ray
instruments, IBIS/ISGRI \citep{u03,lebr03}, on-board
the spacecraft. The effective exposure time is   157 ks and 263 ks for
rev-46 and rev-52, respectively. The data were extracted for all
pointings with a source position offset $\leq$ $12^{\circ}$.
Data reduction was performed using the standard Offline Science
Analysis (OSA) version 3.0 distributed by the {\it INTEGRAL} Science
Data Center \citep{c03}. The algorithms used in the analysis
are described in \citet{gold03}.

\subsection{XMM-Newton}
\label{sec:xmm}

XTE J1807--294 was observed with {\em XMM-Newton}
\citep{j01} on March 22, 2003 for about 17 ks with the first
results reported in \citet{cam03} and \citet{k04}.
The data, simultaneous with the rev-52 {\em INTEGRAL} dataset,
were retrieved from the {\em XMM-Newton} database and re-analyzed
to produce a 0.5--10 keV spectrum contemporary with the {\em INTEGRAL}
observations.
We used the data from the EPIC-pn camera \citep{str01}
operating in Timing mode, the MOS1 camera operating in Small
Window mode and MOS2 in Prime Full Window mode \citep{t01}.
The total amount of good exposure time selected was
14 ks and 9 ks for MOS2 and the other instruments, respectively.
The Observation Data Files (ODFs) were processed to produce calibrated
event lists using  the {\em XMM-Newton} Science Analysis System (SAS
version 5.4.1). The {\em XMM-Newton} data reduction and analysis was
similar to the procedure described in \citet{cam03} with a
special care taken to select the events and eliminate  pile-up.

\subsection{RXTE}
\label{sec:rxte}

We   used publicly available data from the proportional counter array
(PCA) and the High Energy X-ray Timing Experiment
(HEXTE) on board of {\em RXTE}.
XTE J1807--294 was monitored from February  28 to  June 16, 2003.
We analyzed only the data  contemporary with {\em INTEGRAL} for
February 28 to   March 1 and also with {\em XMM-Newton} on March 20--22.
The effective exposure time for the first observation was   15.2
ks and for the later observation  23 ks.
We carried out a spectral analysis, using standard-2 data (with 16 s
time resolution) for the PCA and standard cluster-A data for HEXTE.
To improve the signal-to-noise for the PCA we extracted only the data
from the top layers of PCU2 and PCU3.
The data were extracted for the Good Time Intervals defined
by standard criteria.
The PCA response matrix was created by FTOOLS version 5.3 for 129 energy
channels to cover the energy range from 2--60 keV. For HEXTE we used
the standard 64 energy channel response matrix for the 15--250 keV 
energy range.

\section{Results}
\label{sec:res}

\subsection{IBIS/ISGRI image and light curves}
\begin{figure}
\centerline{\epsfig{file=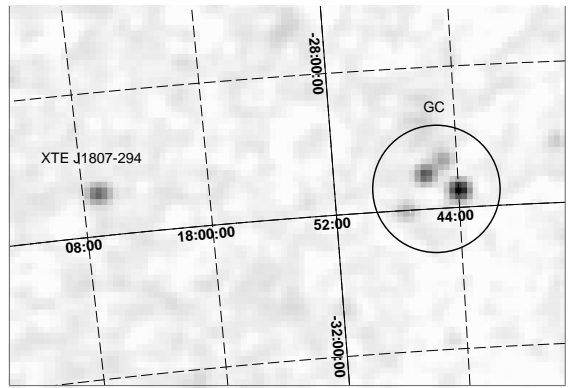,width=7.5cm}}
\caption
{
The {\em INTEGRAL} IBIS/ISGRI mosaiced and deconvolved image in the
20--40 keV energy interval from the rev-46 observation.
Image size is $\sim 9\fdg5 \times 5\fdg5$. The pixel size is
5$'$. The source XTE J1807--294 was detected at a significance of
 28$\sigma$. The other sources  in the Galactic
center region (circle labelled GC) are 1E 1740.7--2942 (53$\sigma$
detection), KS 1741--293 (17$\sigma$), A 1742--294 (33$\sigma$) and
SLX 1744-300 (14$\sigma$).
}
\label{fig:ibis_img}
\end{figure}

Fig. \ref{fig:ibis_img} shows a significance map of the Galactic
center region in the 20--40 keV energy range. Single pointings were
deconvolved and analyzed separately, and then combined in mosaic images.
In the FOV of the image, XTE J1807--294 is clearly detected at a
significance level of $28\sigma$. In the  40--80 keV energy band
  the source was detected at  $14.5\sigma$, and at higher 
energies, 80--120 keV, at $6\sigma$ and from 120--200 keV the detection
significance dropped below $\sim 3\sigma$.
The source was observed at the position
$\alpha_{\rm J2000} = 18^{\rm h}07^{\rm m}00\fs38$ and $\delta_{\rm
  J2000} = -29{\degr}24\arcmin30\farcs8$, with a positional error of
$1\arcmin$ at 90\% confidence \citep{gros03}. This is fully
consistent with the position given by \citet{m03}.
It is located   $5\fdg7$ away from the Galactic center.

\begin{figure}
\centerline{\epsfig{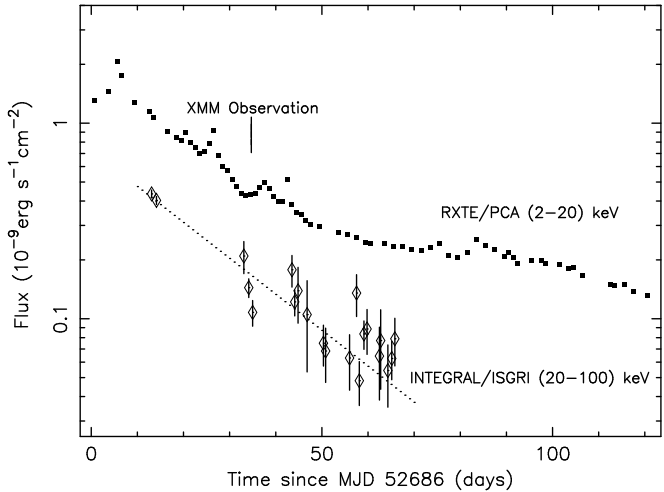}}
\caption
{
{\em RXTE}/PCA and {\em INTEGRAL}/ISGRI light curves in the
  2--20 keV and 20--100 keV  energy band, respectively
(averaged over 1-day intervals).
The {\em INTEGRAL}  data are taken from the GCDE observations
and have been converted to flux units assuming a Comptonization model
(see Table 1). The arrow indicates the {\em XMM-Newton} observation
time.
The dotted line corresponds to $F \propto e^{-t/19.2\ {\rm d}}$ for the
ISGRI data.
}
\label{img:decay}
\end{figure}

The {\em INTEGRAL} 20--100 keV high-energy light curve
has been extracted from the images using all available pointings (all
GCDE observation with the source in the FOV from)
and is shown in Fig.  \ref{img:decay}, averaged over 1-day intervals.
A clear exponential decay ${\rm e}^{-t/\tau}$ with a time-scale of
$\tau = 19\pm 2$ days is seen
over the $\sim$ 60 days of the {\em INTEGRAL} observations.
The corresponding {\em RXTE}/PCA outburst light curve in the 2--20 keV
range is also shown in Fig. \ref{img:decay}. The first five points are
taken from \citet{m03}, where the source was observed
with a peak 2--20 keV flux of $\sim 2 \times 10^{-9}$ erg s$^{-1}$
cm$^{-2}$. The light curve shows also
some short flares during the decay around March 9, 15, 30 and April 5
\citep[see][]{fan05}.
The {\em RXTE} decay time-scale, $\tau = 19.22\pm 0.04$ d, is
fully consistent with the {\em INTEGRAL} value at higher energy.
The equality of the {\em RXTE} and {\em INTEGRAL} timescale shows that
the spectrum did not change much during the outburst. 
After 50 days the decay became slower with a time-scale of
  $122\pm 3$ days.
The XTE J1807--294 outburst lasted much longer than that for other
accretion powered MSPs such as SAX J1808.4--3658 (with $\tau \sim 14$
d) and XTE J1751--305 (with $\tau \sim 7$ d).
XTE J1807--294 is up to now the only MSP which
shows a decay of the outburst profile without a
distinct knee as seen in the other five MSPs \citep[see][]{w05}.

\begin{figure}
\centerline{\epsfig{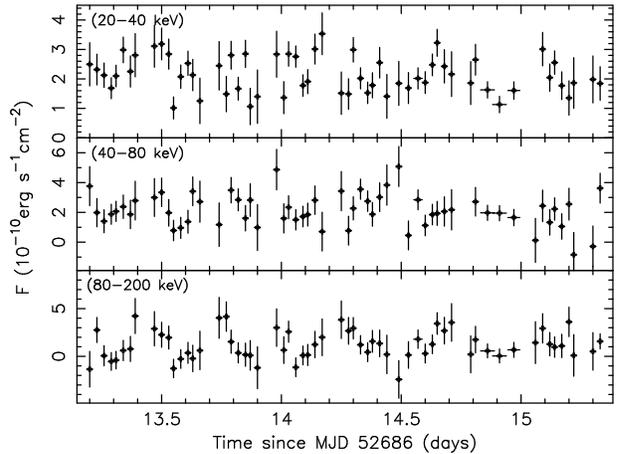}}
\caption
{
The rev-46 high-energy light curve of XTE J1807--294 in different energy
  bands. Each point corresponds to a $\sim2.2$ ks interval.
}
\label{img:lc_ibis}
\end{figure}

Using the rev-46 dataset with the highest S/N, detailed light curves in
different energy bands have been produced, which are shown in Fig.
\ref{img:lc_ibis}.
The count rates, averaged over $\sim2.2$ ks intervals, have been converted
to flux values  using a Comptonization model (see Section 3.2).
The hardness ratio as a function of time
indicates that no significant spectral variability was detected.
The hardness ratio between the highest energy band and the lowest
  energy band was found to be consistent with constant  $0.5\pm 0.1$. 
Pulsations with a period of 5.25 ms were also marginally detected in the
the 20-50 keV IBIS ISGRI energy band. The pulse fraction was $\sim$ 8\%.
These results will be reported in \citet{f05}.

\subsection{Spectral analysis}
\label{sec:analysis}

The spectral analysis was done using XSPEC version 11.3 (Arnaud
1996), combining {\em INTEGRAL}/ISGRI data with available simultaneous
lower energy {\em RXTE} and {\em XMM-Newton} data.
A constant factor was included in the fit to take into account
  the uncertainty in the cross-calibration of the instruments.
A systematic error of 2\% was applied to IBIS/ISGRI spectra which
corresponds to the current uncertainty in the response matrix and 1\%
for all the {\em XMM-Newton} and {\em RXTE} instruments.
All spectral uncertainties in the results are given at a 90\%
confidence level for a single parameter ($\Delta\chi^{2}=2.71$).
The distance, $D$, to the source is not known. As it might be
related to the Galactic center, it is not unreasonable to assume $D
\sim$ 8 kpc. 

\subsubsection{The rev-46 spectrum }
\label{sec:rev46}

In the rev-46 dataset, the joint {\em INTEGRAL}/{\em RXTE} (2.5-200 keV)
spectrum  was first fitted with a simple model consisting
of a photoelectrically-absorbed power-law ({\pl}). Given that we were
not able
to constrain the {\nh} value (as the {\em RXTE}/PCA bandpass starts
above 2.5 keV) we fixed it to the value found from {\em XMM-Newton}
lower energy data (see Section \ref{sec:rev52} below).
A simple {\pl} model is found inadequate with {\chiq}/dof=273/65.
An addition of a cut-off above $\sim$ 60 keV
significantly improves the fit to
{\chiq}/dof=97/64 (probability of chance improvement of $\sim$ 10$^{-14}$).
The best fit is found for a photon index of $\sim$ 1.8 and a
cut-off energy at $\sim$ 80 keV. However, the model does not describe
the spectrum below 15 keV well, which requires a more complex description.

We then substituted this phenomenological model with
a more physical thermal Comptonization model.
The emission from the accretion shock onto the NS
is  well described  by thermal
Comptonization of the black body seed photons from the star
\citep*[see][]{gdb02,pg03,gp04}.
We model the shock emission by the {\sc compps} model\footnote{{\compps}
  is available on the WWW at
  ftp://ftp.astro.su.se/pub/juri/XSPEC/COMPPS} \citep{ps96}.
In this model, the exact numerical solution of the Comptonization
problem  in different geometries is obtained by
considering successive scattering orders.
We approximate the accretion shock geometry by a plane-parallel slab
at the NS surface.
The main model parameters are the Thomson optical depth
  $\tau_{\rm T}$ across the slab, the electron temperature $kT_{\rm e}$,
  and the soft seed photon temperature $kT_{\rm seed}$.
The emitted spectrum depends also on
the angle between the normal and the line of sight $\theta$ which
does not coincide with the inclination of system because of the
light bending.
The seed photons are injected from the bottom of the slab.
A fraction $\exp(-\tau_{\rm T}/\cos \theta)$
of these photons reaches the observer directly,
while the remaining part is scattered in the hot gas.
Thus, the total spectrum contains unscattered black body photons and
a hard Comptonized  tail. 
The best fit with {\chiq}/dof=36/62 is then obtained  for
 $kT_{\rm seed}=$ 0.8 keV,
 $kT_{\rm e} \approx 20$ keV,  $\tau_{\rm T} = 2.7$, and
 $\cos \theta = 0.8$ (see Table 1).
The model also allows to determine the apparent area of the seed
  photons, which turns out to be
$A_{\rm seed}\sim 86 (D/8\ {\rm kpc})^{2}$ km$^2$ in  the best fit. It
corresponds to a  hot spot radius of $\sim5$ km in this outburst phase.

No additional thermal-like component (either blackbody or disk blackbody)
was required by the fit, very likely because the bulk of its
emission occurs below 3 keV,  outside the covered energy range.
Neither iron line emission nor a reflection component were
significantly detected. 

\begin{figure} 
\centerline{\epsfig{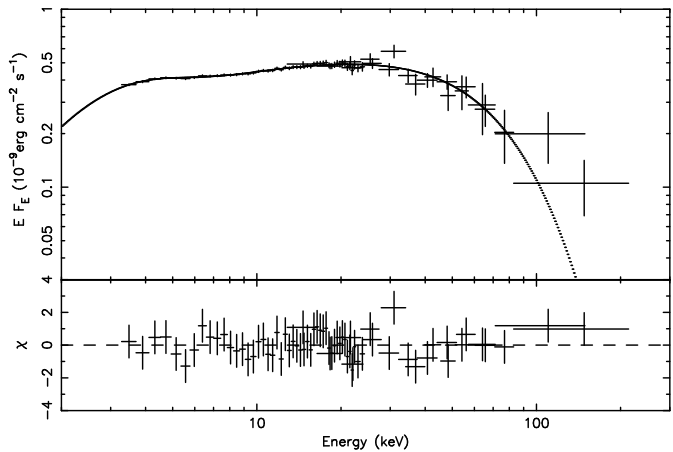}} 
\caption 
{ 
The unfolded spectrum of XTE J1807--294 observed by 
   {\em INTEGRAL}/ISGRI and  {\em RXTE}/PCA/HEXTE during rev-46 along with 
   the best fit {\compps} model. The lower panel shows the residuals between 
   the data and the model. 
} 
\label{fig:spec1} 
\end{figure}

The best fit parameters of the model together with its errors are
reported in Table 1.
The unabsorbed $EF_E$ spectrum and the residuals
of the data to the model are shown in Fig. \ref{fig:spec1}.

\begin{table}[htb]
\begin{center}
\caption{\label{spetab1} Best fit parameters for rev-46 and rev-52 data}
\begin{tabular}{lll}
\hline
\hline
Dataset           & rev-46 & rev-52 \\
Models           & {\compps} & {\compps}+{\sc  diskbb} \\
\hline
$N_{\rm H} (10^{22} {\rm cm}^{-2})$    & $0.56$ (f) & $0.56^{+0.01}_{-0.02}$ \\
$kT_{\rm disk}$   (keV)                & -          & $0.43^{+0.04}_{-0.04}$ \\
$R_{\rm in}\sqrt{\cos \, i}$ (km)  & -          & $13.4^{+2.2}_{-1.3}$ \\
$kT_{\rm e}$ (keV)                  & $18^{+3}_{-3}$    & $37^{+28}_{-10}$ \\
$kT_{\rm seed}$ (keV)    & $0.8^{+0.04}_{-0.05}$  & $0.75^{+0.04}_{-0.04}$ \\
$\tau_{\rm T}$           & $2.7^{+0.1}_{-0.2}$    & $1.7^{+0.5}_{-0.8}$ \\
$A_{\rm seed}$ (km$^2$)  & $86^{+15}_{-8}$  & $26^{+23}_{-12}$ \\
$\cos \, \theta $         & $0.80^{+0.06}_{-0.09}$  & $0.79^{+0.07}_{-0.06}$ \\
$\chi^{2}/{\rm dof}$                   & $36/62$     & $214/198$ \\
$L_{\rm 0.1-200 keV}$  ($10^{37}$ erg s$^{-1}$) & $1.26$        & $0.36$  \\
\hline
\end{tabular}
$^a$ Assuming a distance of 8 kpc.
\end{center}
\end{table}

\subsubsection{The rev-52 spectrum}
\label{sec:rev52}

The {\em INTEGRAL} rev-52 dataset was obtained 22 days later than rev-46,
this time also simultaneously with the {\em XMM-Newton} ToO observation.
We were therefore able to construct a combined spectrum covering the
0.5--200 keV range.
The thermal emission from the accretion disk can contribute
at low energies in the {\em XMM-Newton} range.
The presence of the emission from the accretion disk
is supported by the decrease of the pulse fraction below 2 keV
\citep{cam03,k04}.
We model this emission with {\sc  diskbb},
a multi-temperature disk model of \citet{m84}.
It  is characterized by the inner disk
temperature $kT_{\rm in}$ and normalization which can be transformed to the
inner disk radius $R_{\rm in}$.
The spectrum was then fitted with this
 component combined with a simple {\pl}, everything photoelectrically
 absorbed. The model gives a good fit with {\chiq}/dof=237/200.
The absorption is determined to be $N_{\rm H} = 0.56 \times 10^{22}
{\rm cm}^{-2}$, comparable to that found by \citet{cam03}, and the best
{\pl} index is  $\Gamma \sim 1.9$. 
However, the inner disk radius, 
$R_{\rm in}\sqrt{\cos \, i}$, is then around 2 km, which is not consistent 
with the disk size, assuming a 10 km NS radius. 
To have a direct comparison with the results of the rev-46 
data we replaced the {\pl}   component with {\compps} 
in a slab geometry  assuming a  black body seed photon spectrum. 
 
\begin{figure} 
\centerline{\epsfig{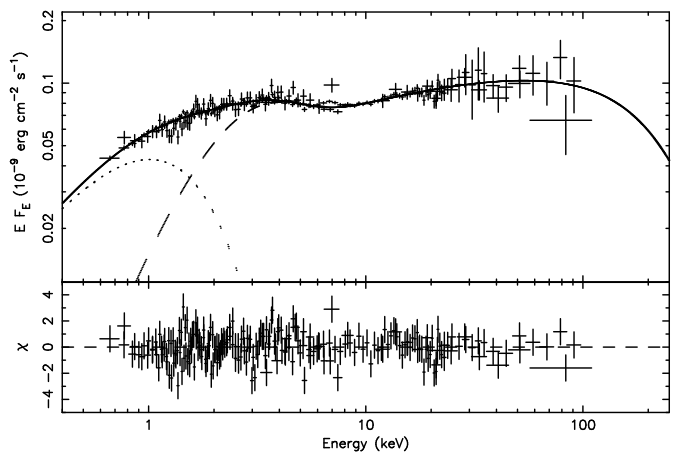}} 
\caption 
{ 
Simultaneous {\em INTEGRAL}, {\em XMM-Newton} and {\em 
      RXTE}  spectra of XTE 
      J1807--294 fitted with an absorbed disk black body, {\sc
	diskbb}, plus {\sc compps}  
model. The EPIC-pn and MOS2 spectra in the 
0.5--10 keV range and PCA/HEXTE in the 3--200 keV and IBIS/ISGRI in the 
      20--200 keV range are shown. The {\dbb} model is shown by a dotted curve, 
the dashed curve gives the {\compps} model, while the total spectrum is 
      shown by a solid curve. The lower panel presents the residuals. 
} 
\label{fig:spec2} 
\end{figure}

The results of the {\sc  diskbb} + {\compps} fit are reported in Table
1, while the $EF_E$ spectrum and the residuals  are shown in
Fig.~\ref{fig:spec2}. With a typical plasma temperature of $kT_{\rm e}
\sim 40$ keV, the joint {\em INTEGRAL/XMM-Newton/RXTE} fit is at variance 
with the parameters derived from the low energy {\em XMM-Newton} data 
only  which gave  $kT_{\rm e} \sim$ 10 keV \citep[see][]{cam03},
confirming the importance of high energy observations. 
Comparison with the rev-46 results shows that the decay of the outburst is 
marked by a constant seed photon temperature, a 
significant increase of plasma temperatures correlated with a 
significant decrease of the scattering opacity. 
The fit required a inner disk radius, $R_{\rm in}\sqrt{\cos\
  i} \sim$  
13 km, this time fully compatible with the radius of a NS. 
The apparent area of the seed photons is $A_{\rm seed} \sim 26 (D/8 
\ {\rm kpc})^{2} {\rm km}^{2}$, which is also what expected from a hot spot on 
a NS surface. 
The hard component contributes 83 per cent of the unabsorbed 
flux in the 0.1-200 keV energy range.
Also in this case reflection was not significantly detected. 

\section{Companion star and the system geometry} 
 
\begin{figure} 
\centerline{\epsfig{file=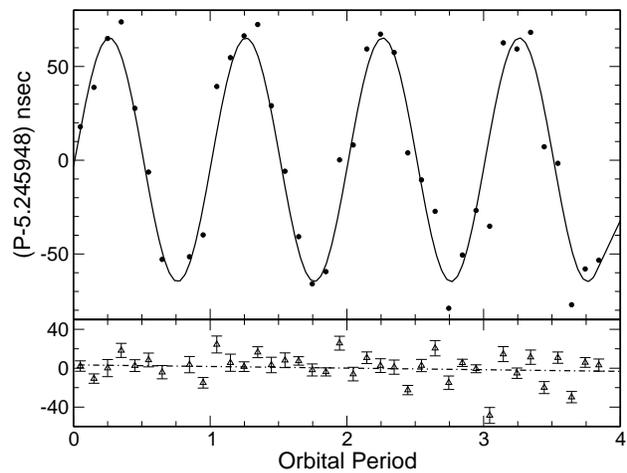,width=7.2cm,angle= -90}} 
\vspace{-0.4 cm} 
\caption 
{ 
Doppler variations of the spin period as a function of orbital phase 
and the best sine fit. 
Each point corresponds to a 240 s interval. 
Significant residuals are seen with rms of $\sim 15$ ns. 
} 
\label{fig:dopres} 
\end{figure}

The orbital elements of XTE J1807--294 have been derived from a 
re-analysis of the {\em XMM-Newton} temporal data, using a Doppler 
pulse-variation procedure instead of the complex 2D-parameter search 
made by \citet{cam03} and \citet{k04}. 
The barycentric photon arrival times were grouped into 240 s intervals 
and the best period 
was determined using the Z-statistic  \citep{b83}. 
The resulting 39 best period values clearly show a sinusoidal 
modulation expected for a circular orbit (see Fig. \ref{fig:dopres}). A 
sine-fit to the modulation yields a mean pulsation period of
0.005245948(4) s, a  
projected orbital radius of $a_{\rm x}\sin {\rm i} = 4.75(39)\times
10^{-3}$ lt-s and a barycentric positive-crossing time (mean longitude
$+ 90^{\circ}$) of $T_{0}= 2452721.168(2)$ JD, 
where the errors are given in  parentheses. 
The poorly constrained orbital period was assumed here to 
be 40.0741 min  \citep{mss04}. The spin period and orbit 
projected radius values are consistent with the previous 
determinations by \citet{cam03} (once corrected for typing 
errors) and \citet{k04}, but with more reliable 
uncertainties. However, the epoch of mean longitude differs very
significantly,  
by more than 0.005 d, most probably due to the inaccuracy of the complex 
method used by these authors. The introduction of an eccentricity does
not improve the fit.  
A small linear slope is visible in the residuals (see
Fig. \ref{fig:dopres}), but not significant. However phase deviations
(rms of $\sim 15$ ns) are also seen (Fig. \ref{fig:dopres}), 
corresponding to highly erratic period variations. 
Such a behavior was already 
noted by \citet{m04} who reported apparent spin-up and spin-down 
rate of $\sim \pm 3 \times 10^{-16}$ s s$^{-1}$ on longer timescale. The 
instability of the period may therefore be also present on very short 
timescale ($\sim 5$ min) which is a strong indication that we do not 
see the real clock, 
but rather a moving spot at the surface of the neutron star. 

With a mass function of $f(m) = 1.49^{+0.40}_{-0.34} \times 10^{-7}$ 
M$_{\odot}$, the minimum companion 
mass obtained for inclination of 
$i = 90\degr$ is  $M_{\rm c} = 0.0053, 0.0066$ and 
0.0084 M$_{\odot}$ for $M_{\rm x} = 1.0, 1.4$ and 2.0 M$_{\odot}$,
respectively. 

\begin{figure} 
\centerline{\epsfig{file=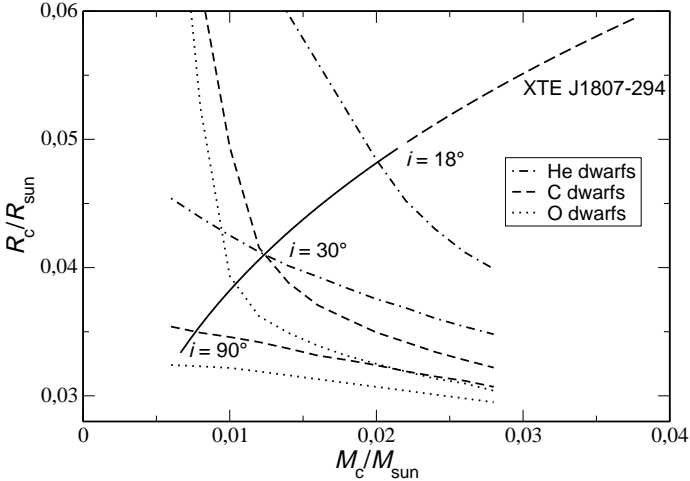,width=7.8cm,angle= -90}} 
\caption 
{ 
Companion mass $M_{\rm c}$ vs. radius $R_{\rm c}$ plane, showing 
the Roche lobe constraints for XTE J1807--294 (thick line), for 
$M_{\rm x}=1.4 {\rm M}_{\odot}$. 
The figure shows also   the low mass regime degenerate dwarf 
models incorporating different compositions (dotted O, dashed C, 
dot-dash He) and low ($10^{4}$ K) or high ($3 \times 10^{6}$ K) 
central temperatures (lower and upper curves). 
} 
\label{fig:rlob} 
\end{figure}

The assumption of a Roche lobe filling companion \citep{p71}, 
implies the mass-radius relation shown in Fig. \ref{fig:rlob} (thick line). 
In the very low-mass regime, recent models of low-mass degenerate 
dwarfs have been produced incorporating the effect of different 
compositions and temperatures \citep{db03}. The 
corresponding $M_{\rm c}$ versus $R_{\rm c}$ relations are also shown in 
Fig. \ref{fig:rlob}. 
For a mean inclination of $i\sim 60^{\circ}$, only low temperature C 
dwarfs and   O dwarfs are allowed, clearly excluding He dwarfs 
which will require very low inclinations $i\leq 30\degr$. 
This  strongly constrains   the evolution of the system. 
 
An interesting constraint is given by the minimum
accretion rate driven by  gravitational radiation in a close binary
\citep[see][for a review]{vh95}. The outburst main peak was observed on
 February 21, 2003 (52691 MJD), and follows the
outburst shape (see Fig. \ref{img:decay}), we extimated a total flux
to be $7.7\times\, 10^{-3}$ erg cm$^{-2}$.  Following
\citet{bc01} we can
constrain the source distance as a function of inclination angle and
companion mass through the mass function and mass loss rate due to
gravitational rediation.  From randomly selected orbits, the
probability of having an inclination $i < i_{\rm \circ}$ is
1-cos$i_{\rm \circ}$, so that a maximum mass at a confidence level 95\%
 can be derived for $i < 18^{\circ}$ with value $M_{\rm c} \leq
 0.022{\rm M}_{\odot}$. Note that for this source the recurrence time
of the outburst is unknown. Assuming a recurrence time of 5 years, the source
is calculated to be at 8 kpc with an inclination angle of $18^{\circ}$
($M_{\rm c} =0.022{\rm M}_{\odot}$). At a mean inclination
angle of $60^{\circ}$ ($M_{\rm c} =0.0077{\rm  M}_{\odot}$), the source
must be located at 3 kpc. With a recurence time of 39 years, the
source is located at 8 kpc with an inclination angle of $60^{\circ}$.
This calculation predicts that a He dwarf companion star is favorable
for a low inclination angle and the system located at $\sim$8
kpc. A high inclination implies that the system  be located at $\sim$3
kpc with an C/O dwarf companion star. A solution of this problem can
be given if a future outburst is observed, giving  a recurrence time
of outburst. 

The mass transfer driven by gravitational  radiation implies an
inclination $i>60\degr$, assuming a source distance of 8 kpc and
NS masses between 1.0 and 2.0 M$_{\odot}$, giving only a long recurence
time of the outburst. On the other hand, there were no X-ray eclipses
or dips detected in the EPIC-pn light curves. We therefore put an
upper limit on the binary inclination of $i<83\degr$ for a Roche lobe-filling
companion.
A very low inclination is also possible, since that would imply a high
companion mass and thus a large mass transfer rate.
There is also no evidence for an X-ray modulation at the
binary period, which might have implied propagation through a
scattering atmosphere in a near edge-on geometry \citep{bc01}.
If the source is most likely at an inclination of $60\degr-83\degr$,
we get the inner disk radius (see Table 1) in the range 20--40 km.

\section{Conclusions} 
 
We have found that the spectrum of XTE J1807--294 is well described by 
a combination of thermal Comptonization  and  disk black body. 
The electron temperature of $kT_{\rm e}\sim$ 40 keV 
 is  considerably larger than previously determined from 
lower energy observations \citep{cam03}. The hard spectral 
component contributes most of the observed flux (83 per cent), even though 
a soft component (disk black body) is required by the data. 
If the accretion disk is truncated at the 
magnetosphere radius, where the material can accrete along the magnetic 
field towards the pole of a NS, the matter over the 
pole is heated by a shock to a temperature of $\sim$ 40 keV. The hot spot 
at the surface with temperature $\sim$ 0.8 keV gives rise to the seed 
photons for Comptonization in the hot plasma. Since the hard X-ray 
emission is pulsed \citep*[see e.g.][]{gp04}, a fraction of it must 
originate from the regions confined by the magnetic field. The most 
obvious source of hard X-rays is the place where material collimated by 
the magnetic fields impacts  the surface. 
 
If we consider the inclination of the system to be $60\degr<i<83\degr$, 
this allows us to determine the inner disk radius which lies 
in the range 20--40 km (for the distance of 8 kpc). 
Many characteristics of the source are 
similar to those observed in other MSPs, SAX J1808.4--3658 \citep{gdb02,pg03} 
and XTE J1751--305 \citep{gp04}.

\acknowledgements 
MF and AM acknowledges CNES and the CNRS ``GdR Phenomenes Cosmiques de Haute 
Energie'' for financial support. JP is supported by the Academy of Finland and 
the NORDITA Nordic project on High Energy Astrophysics.

\end{document}